\documentstyle[12pt]{article}

\newcommand{\beq}{\begin{equation}}
\newcommand{\eeq}{\end{equation}}
\newcommand{\beqa}{\begin{eqnarray}}
\newcommand{\eeqa}{\end{eqnarray}}

\newcommand{\eq}[1]{(\ref{#1})}

\newcommand{\ra}{\rightarrow}

\newcommand{\NP}[1]{ {\it Nucl.~Phys.} {\bf #1}}
\newcommand{\PL}[1]{ {\it Phys.~Lett.} {\bf #1}}

\newcommand{\PR}[1]{ {\it Phys.~Rev.} {\bf #1}}
\newcommand{\PRL}[1]{ {\it Phys.~Rev.~Lett.} {\bf #1}}

\begin{document}
\topmargin 0pt
\oddsidemargin 1mm
\begin{titlepage}
\begin{flushright}
NBI-HE-95-40\\
hep-th/9511158\\
\end{flushright}
\setcounter{page}{0}
\vspace{15mm}
\begin{center}
{\Large  Relation between worldline Green functions 
         \\for scalar two-loop diagrams } 
\vspace{20mm}

{\large Haru-Tada Sato   
\footnote{ Fellow of the Danish Research Academy,\,\,\, 
 sato@nbivax.nbi.dk}}\\
{\em The Niels Bohr Institute, University of Copenhagen\\
     Blegdamsvej 17, DK-2100 Copenhagen, Denmark}\\
\end{center}
\vspace{7mm}

\begin{abstract}
We discuss a relation between two-loop bosonic worldline Green 
functions which are obtained by Schmidt and Schubert in two different 
parametrizations of a two-loop worldline. These Green functions are 
transformed into each other by some transformation rules based on 
reparametrizations of the proper time and worldline modular parameters. 
\end{abstract}

\vspace{1cm}

\end{titlepage}
\newpage
\renewcommand{\thefootnote}{\arabic{footnote}}
\indent
 One of interesting and important theoretical aspects of quantum field 
theories, in particular of gauge theories, is the Bern-Kosower 
formalism \cite{BK} which provides a reformulation of one-loop Feynman 
amplitudes. Their idea is very natural that amplitudes in ordinary field 
theory may be reproduced by the infinite string tension limit of superstring 
amplitudes. It results in a new set of rules instead of Feynman rules and 
enables several calculations: five point gluon \cite{gl5}, four point 
graviton amplitudes \cite{gr4} and so on \cite{others}. The advantage 
of this formalism is to get rid of handling a vast number of Feynman 
diagrams and of taking care of gauge cancelation of diagrams in gauge 
invariant theories. It might clarify underlying structure, which we have not 
been recognized yet in gauge theories; for example, various useful 
informations have been developed in worldline approaches 
\cite{info}-\cite{axial}.
         
Extension of this formalism to multi-loop diagrams has also been 
investigated recently \cite{SSphi},\cite{SSqed},\cite{sumino} generalizing 
Strassler's approach \cite{S} to the Bern-Kosower rules. Strassler derived 
the worldline Green functions \cite{poly} of spinor, scalar and gauge fields 
for one-loop diagrams rewriting one-loop effective actions as path integrals 
of (supersymmetric) worldline actions. In one-loop case, the modular 
parametrization of a loop is unique and we can define the worldline Green 
functions in a unique way. However, this situation changes in multi-loop 
cases because we have a variety of choices of the parametrization due to node 
points, i.e., internal vertices. Schmidt and Schubert have actually 
obtained two expressions for the two-loop worldline Green function in the 
$\phi^3$ theory \cite{SSphi}. They explicitly checked that amplitudes 
derived from these respective Green functions coincide with at least three 
or four point functions from the Feynman rule calculations.
                             
It is apparent that this kind of equivalence check between worldline Green 
functions in different forms becomes difficult in more general situations 
such as higher loop diagrams and a large number of external legs. In addition, 
we can not recognize which parametrization is natural or convenient one in a 
complicated case, and hence it is useful to know how to convert the Green 
functions to a differently parametrized form. It is therefore important to 
find a clear connection between multi-loop Green functions defined on a 
differently parametrized worldline. 

In this report, we discuss a relationship between two-loop bosonic worldline 
Green functions proposed by Schmidt and Schubert in the scalar $\phi^3$ 
theory. First let us survey the vacuum amplitude at one-loop level in two 
different ways which are instructive to get an insight into two-loop case. 
One is given by the path integral for a worldline action with cyclic boundary 
condition and the other may be given by sewing two propagators of path 
integral expression \cite{S},\cite{poly};
\beq
I_0^{(1)}=\int_0^{\infty}{dT\over T}N(T)\int_{x(0)=x(T)}{\cal D}x
\exp[-\int_0^T d\tau {1\over4}{\dot x}^2(\tau)]
=\int{dT\over T}(4\pi T)^{-D/2}         \label{no1}                        
\eeq   
and 
\beq
J_0^{(1)}=\int d^Dx_a d^Dx_b \prod_{i=1}^2 \int_0^{\infty} dT_i 
        {\tilde N}_i(T_i) \int_{\scriptstyle x_i(0)=x_a 
        \atop\scriptstyle x_i(T_i)=x_b}{\cal D}x_i
       \exp[-\int_0^{T_i} d\tau {1\over4}{\dot x}_i^2(\tau)] \label{no2}
\eeq\[
  =\int d^Dx_b \int dT_1 dT_2 [4\pi (T_1+T_2)]^{-D/2}, 
\]                                                    
where the path integral normalizations $N$ and ${\tilde N}$ are determined 
as the second equality in each case. We omit the mass term $e^{-m^2T}$ 
in \eq{no1} for simplicity. Comparing these equations, we think of the 
transformation from $J_0^{(1)}$ to $I_0^{(1)}$ in the following form
\beq
              T_1 =T(1-u),\quad T_2=Tu                \label{no3}
\eeq
which gives
\beq
     \int_0^{\infty}dT_1 dT_2 = \int_0^{\infty}dTT\int_0^1 du.  \label{no4}
\eeq
RHS of the latter expression \eq{no2} becomes  
\beq
\int d^D x_b \cdot \int_0^{\infty}{dT\over T}T^2\int_0^1 du(4\pi T)^{-D/2}
=(2\pi)^D\delta^D(0)\cdot\int_0^{\infty}{dT\over T}
 \int_0^T d\tau_a \int_0^T d\tau_b (4\pi T)^{-D/2},         \label{no5}
\eeq
where we have used the fact that the integrand is independent of any $\tau$ 
variables. If we consider $N$-point functions as discussed later, we see 
that $(2\pi)^D\delta^D(0)$ corresponds to the conservation law of external 
momenta, which is implicit in \eq{no1}. The integrals with respect to $\tau_a$ 
and $\tau_b$ correspond to node point integrals which appear in multi-loop 
integrals. Here we inevitably encounter the node point integrals because we 
need an additional leg at a glue point of two propagators regardless of 
external or internal leg. Strictly speaking, eq.\eq{no5} is not so much a 
one-loop vacuum amplitude as a two-point function with external zero momenta 
or a kind of two-loop vacuum amplitude of one internal line of infinite length 
(this will become clear later). We hence pick up pure one-loop 
contribution from \eq{no5} through dropping node point integrals by hand. In 
this way, we understand that \eq{no3} relates \eq{no1} with \eq{no2} which is 
symmetrized in $T_1$ and $T_2$.

We further proceed to $N$-point functions which concern the worldline Green 
functions. Evaluating path integrals with insertion of $N$ scalar vertex 
operators in both \eq{no1} and \eq{no2}, we have the following two 
expressions which are seemingly different from each other 
\beq
I_N^{(1)}=\int_0^{\infty}{dT\over T}(4\pi T)^{-D/2}\cdot
\prod_{n=1}^N \int_0^T d\tau_n \exp[{1\over2}\sum_{j,k}^N
                p_j p_k G_B(\tau_j,\tau_k)]            \label{no6}
\eeq 
and 
\[
J_N^{(1)}=(2\pi)^D\delta^D(\sum_{n,i} p^{(i)}_n)\int_0^{\infty}dT_1 dT_2 
          [4\pi (T_1+T_2)]^{-D/2}\cdot
          \prod_{n=1}^{N_1} \int_0^{T_1} d\tau^{(1)}_n 
          \prod_{n=1}^{N_2} \int_0^{T_2} d\tau^{(2)}_n \]\beq 
\times \exp[{1\over2}\sum_{a=1}^2\sum_{j,k}^{N_a} p^{(a)}_j p^{(a)}_k 
                          G_{aa}(\tau^{(a)}_j,\tau^{(a)}_k)
             +\sum_j^{N_1}\sum_k^{N_2} p^{(1)}_j p^{(2)}_k 
                          G_{12}(\tau^{(1)}_j,\tau^{(2)}_k) ] \label{no7}
\eeq  
where $G_B$ is the one-loop bosonic Green function \cite{S},\cite{poly}
\beq
        G_B(x,y)=\vert x-y\vert -{(x-y)^2\over T}        \label{no8}
\eeq
and $G_{ij}$ are 
\beq
G_{11}(x,y)=G_{22}(x,y)=\vert x-y\vert - {(x-y)^2\over T_1+T_2} \label{no9}
\eeq
\beq
          G_{12}(x,y)=x+y - {(x+y)^2 \over T_1+T_2}.         \label{no10}
\eeq
In \eq{no9} and \eq{no10}, $G_{ij}$ are rearranged from original path 
integral results with the use of momentum conservation in accordance with 
two-loop results \cite{SSphi}. Note that $\tau^{(2)}$ is defined to run in the 
opposite direction to $\tau$ and $\tau^{(1)}$ because we have chosen the 
boundary condition of path integral \eq{no2} as $x_i(0)=x_a$ and 
$x_i(T_i)=x_b$ for $i=1,2$. If we choose $x_1(0)=x_2(T_2)=x_a$ and 
$x_1(T_1)=x_2(0)=x_b$, we will obtain an expression with reversed sign for 
$\tau^{(2)}$ variable. Since we have seen a correspondence between 
$T$-parameters' integral measures in \eq{no5}, we do not have to repeat that 
here. Changing the sign of $\tau^{(2)}$ and renaming the variables 
$\{\tau^{(1)}_n,\tau^{(2)}_n\}$ and $\{p^{(1)}_n,p^{(2)}_n\}$ as 
$\tau_n$ and $p_n$, we obtain the same equation as \eq{no6} after omitting 
the delta function and node point integrals as a result of the transformation 
\eq{no3}. Here we have assumed that the transformation of $\tau$-integration 
measures follow a naive replacement
\beq
\prod_{n=1}^N \int_0^T d\tau_n \quad\ra\quad 
\prod_{i=1,2}\prod_{n=1}^{N_i} \int_0^{T_i} d\tau^{(i)}_n.   \label{no11}
\eeq

Let us promote the above one-loop idea to the two-loop case. 
Two-loop $N$-point functions with different parametrizations from each other 
are obtained through insertion of one propagator parametrized by length 
$T_3$ 
\beq
    \int_0^{\infty} dT_3 {\tilde N}_3(T_3) 
    \int_{\scriptstyle x_3(0)=x_a \atop\scriptstyle x_3(T_3)=x_b}{\cal D}x_3
    \exp[-\int_0^{T_3} d\tau {1\over4}{\dot x}_3^2(\tau)]. \label{no12}
\eeq
The insertion of this propagator and scalar vertex operators into \eq{no1} 
and \eq{no2} leads to \cite{SSphi}
\[
I_N^{(2)}=\int_0^{\infty}{dT\over T}\int_0^{\infty}dT_3 (4\pi)^{-D}
\int_0^T d\tau_a \int_0^T d\tau_b [TT_3+TG_B(\tau_a,\tau_b)]^{-D/2}
\prod_{n=1}^N \int_0^T d\tau_n \]\beq
\times\exp[{1\over2}\sum_{j,k}^N
          p_j p_k G^{(1)}_B(\tau_j,\tau_k)]   \label{no13}
\eeq  
and 
\[
J_N^{(2)}= \prod_{a=1}^3\int_0^{\infty}dT_a(4\pi)^{-D}
(T_1T_2+T_2T_3+T_3T_1)^{-D/2} 
          \prod_{n=1}^{N_1} \int_0^{T_1} d\tau^{(1)}_n 
          \prod_{n=1}^{N_2} \int_0^{T_2} d\tau^{(2)}_n \]\beq 
\times \exp[{1\over2}\sum_{a=1}^2 \sum_{j,k}^{N_a} p^{(a)}_j p^{(a)}_k 
                      G^{\rm sym}_{aa}(\tau^{(a)}_j,\tau^{(a)}_k)
        +\sum_j^{N_1}\sum_k^{N_2} p^{(1)}_j p^{(2)}_k 
                 G^{\rm sym}_{12}(\tau^{(1)}_j,\tau^{(2)}_k) ], \label{no14}
\eeq
where our $G^{(1)}_B$ and $G^{\rm sym}_{ij}$ are same ones in \cite{SSphi} 
denoted by ${\tilde G}^{(1)}_B$ and ${\tilde G}^{\rm sym}_{ij}$; namely, 
\beq
G^{(1)}_B(x,y)=G_B(x,y)-{1\over4}{ 
  \left( G_B(x,\tau_a)-G_B(x,\tau_b)-G_B(y,\tau_a)+G_B(y,\tau_b)\right)^2
      \over T_3+G_B(\tau_a,\tau_b)}             \label{no15} 
\eeq
and
\beq
G^{\rm sym}_{aa}(x,y)=\vert x-y\vert -{T_{a+1}+T_{a+2}\over 
                   T_1T_2+T_2T_3+T_3T_1}(x-y)^2            \label{no16}
\eeq
\beq
G^{\rm sym}_{aa+1}(x,y)=x+y-{x^2T_{a+1}+y^2T_{a}+(x+y)^2T_{a+2} \over
                      T_1T_2+T_2T_3+T_3T_1}.            \label{no17}
\eeq
The subscripts on $G$ mean the superscripts of first and second arguments 
as easily understood in \eq{no14} and $T_4\equiv T_1$. 
It is worth noticing that the integrand of \eq{no14} coincides with that of 
\eq{no7} in the limit $T_3\ra\infty$. This implies the previous note 
that $J^{(1)}_0$ may be regarded as a two-loop contribution from an infinite 
internal line. 

Similarly to the one-loop case, we begin with the vacuum ($N=0$) case. 
According to the change of variables \eq{no3} and \eq{no4}, eq.\eq{no14} 
becomes
\beq
J_0^{(2)}= \int_0^{\infty}dT_3 dTT\int_0^1du(4\pi)^{-D}
           [TT_3+T^2u(1-u)]^{-D/2}.                    \label{no18}
\eeq
We then transform the variable $u$ in order to reproduce the Green 
function $G_B(\tau_a,\tau_b)$ which can be seen in \eq{no13}
\beq
        u={\vert \tau_a -\tau_b\vert \over T}.        \label{no19}
\eeq
Taking account of this transformation, eq.\eq{no18} coincides 
with the vacuum amplitude $I_0^{(2)}$
\beq
J_0^{(2)}=I_0^{(2)}
         =\int_0^\infty dT_3{dT\over T}\int_0^Td\tau_a\int_0^Td\tau_b 
          (4\pi)^{-D}[TT_3+TG_B(\tau_a,\tau_b)]^{-D/2}.         \label{no20}
\eeq 
We here observe that \eq{no19} is needed in addition to \eq{no3} differently 
from one-loop case \eq{no5} where $u$ disappears.

In the case of $N$-point functions, we need a further consideration. 
We use the fact that to deal only with $\tau_a\leq\tau_b$ is guaranteed by 
the symmetry of \eq{no13} under $\tau_a\leftrightarrow\tau_b$. The reversal 
of $\tau^{(2)}$ considered in the one-loop case should be modified taking 
account of dependence on the proper time position of a node point;
\beq
    \tau_n^{(1)}=x_n-\tau_b,\quad \tau_n^{(2)}=\tau_b-y_n.  \label{no21}
\eeq
This is the $\tau^{(2)}$ reversal transformation with respect to $\tau_b$ 
which is regarded as the origin of $\tau^{(1)}$ coordinate shifted by $\tau_b$ 
accordingly. In one-loop case, we can put the origin $\tau_b=0$ in safety 
(because of no $u$ dependence). Notice also that to adopt $\tau_a$ as the 
origin of the reversal transformation instead of $\tau_b$ corresponds to 
consider $\tau_b\leq\tau_a$ exchanging $\tau_a\leftrightarrow\tau_b$. 
From \eq{no3} and $0\leq\tau^{(i)}\leq T_i$, we see also
\beq
       \tau_a \leq y_n\leq \tau_b \leq x_n.       \label{no22}
\eeq
It is convenient to write down $G^{(1)}_B$ explicitly according to this 
sequence \eq{no22},
\[
G^{(1)}_B(x_n,x_m)=G_B(x_n,x_m)-{\vert\tau_a-\tau_b\vert^2\over 
         T_3+G_B(\tau_a,\tau_b) }{(x_n-x_m)^2\over T^2} 
\]\beq
G^{(1)}_B(y_n,y_m)=G_B(y_n,y_m)-{(T^2-\vert\tau_a-\tau_b\vert^2)^2\over 
         T_3+G_B(\tau_a,\tau_b) }{(y_n-y_m)^2\over T^2}   \label{no23}
\eeq\[ 
G^{(1)}_B(x,y)=G_B(x,y)-{1\over T_3+G_B(\tau_a,\tau_b)} 
(\tau_b-y-\vert\tau_a-\tau_b\vert{x-y\over T})^2.
\]
With substitution of \eq{no3},\eq{no19},\eq{no21} in \eq{no16}, we verify 
that $G^{\rm sym}_{ij}$ for $i,j=1,2$ are transformed into $G^{(1)}_B$ as 
follows:
\[
G^{\rm sym}_{11}(\tau^{(1)}_n,\tau^{(1)}_m)=G_B^{(1)}(x_n,x_m),
\]\beq
G^{\rm sym}_{22}(\tau^{(2)}_n,\tau^{(2)}_m)=G_B^{(1)}(y_n,y_m), \label{no24}
\eeq\[
G^{\rm sym}_{12}(\tau^{(1)}_n,\tau^{(2)}_m)=G_B^{(1)}(x_n,y_m).
\]

Finally, we apply our transformation to another set of worldline Green 
functions which can be obtained from inserting vertex operators also into the 
$T_3$ propagator part \eq{no12}. The exponential parts including Green 
functions of \eq{no13} and \eq{no14} are generalized to the following 
respectively 
\beq
\exp[{1\over2}\sum_{jk}^{N'} p_j p_k G^{(1)}_{11}(\tau_j,\tau_k)
     +{1\over2}\sum_{jk}^{N_3} p^{(3)}_j p^{(3)}_k 
                    G^{(1)}_{33}(\tau^{(3)}_j,\tau^{(3)}_k)
     +\sum_j^{N'}\sum_k^{N_3} p_j p^{(3)}_k 
                    G^{(1)}_{13}(\tau_j, \tau^{(3)}_k) ]   \label{no25}
\eeq
and
\beq 
\exp[{1\over2}\sum_{a=1}^3 \sum_{j,k}^{N_a} p^{(a)}_j p^{(a)}_k 
                 G^{\rm sym}_{aa}(\tau^{(a)}_j,\tau^{(a)}_k)
        +\sum_{a=1}^3\sum_j^{N_a}\sum_k^{N_{a+1}} p^{(a)}_j p^{(a+1)}_k 
              G^{\rm sym}_{aa+1}(\tau^{(a)}_j,\tau^{(a+1)}_k) ], \label{no26}
\eeq
where $N'=N_1+N_2$, $N_4=N_1$ etc., and $G^{(1)}_{ij}$ are
\beq
G^{(1)}_{11}(x,y)=G^{(1)}_B(x,y),     \label{no27}
\eeq\beq
G^{(1)}_{33}(z_1,z_2)=\vert z_1-z_2\vert
              -{(z_1-z_2)^2\over T_3+G_B(\tau_a,\tau_b)},  \label{no28}
\eeq\beq
G^{(1)}_{13}(x,z)=G_B^{(1)}(x,\tau_a)+{1\over T_3+G_B(\tau_a,\tau_b)}
\left(T_3z- z^2+ z[G_B(x,\tau_b)-G_B(x,\tau_a)]\right)      \label{no29}
\eeq                                                              
or
\beq
G^{(1)}_{13}(x,z)=G_B^{(1)}(x,\tau_b)+{1\over T_3+G_B(\tau_a,\tau_b)}
\left(T_3z- z^2 + z[G_B(x,\tau_a)-G_B(x,\tau_b)]\right).      \label{no30}
\eeq                                                              
Eqs.\eq{no29} and \eq{no30} are different from each other, however they are 
essentially same under the exchange of integral variables 
$\tau_a\leftrightarrow\tau_b$. If we follow the rule of transformations 
\eq{no3},\eq{no21} and \eq{no22} for $\tau_a\leq\tau_b$, we have to choose 
\eq{no30} which can be derived from boundary conditions $x(0)=x(\tau_b)$ and 
$x( T_3)=x(\tau_a)$ in the path integral \eq{no12}, where $\tau_b$ is regarded 
as the origin of $\tau$ parameter. If we want to discuss \eq{no29}, we have 
to use the transformations (obtained from $\tau_a\leftrightarrow\tau_b$) 
for the reverse case $\tau_b\leq\tau_a$ as mentioned above. This situation is 
completely same as the third equation in \eq{no23}. As for \eq{no27} 
and \eq{no28}, we have no attention to such choice as whether \eq{no29} or 
\eq{no30} because they are already symmetric form in $\tau_a$ and $\tau_b$. 

Eq.\eq{no27} has been checked in \eq{no24} to be transformed into 
$G^{\rm sym}_{11}$, $G^{\rm sym}_{22}$ and $G^{\rm sym}_{12}$. 
Eq.\eq{no28} also coincides with $G^{\rm sym}_{33}$ under our transformation 
rules discussed above
\beq
G^{(1)}_{33}(\tau^{(3)}_n,\tau^{(3)}_m)=
G^{\rm sym}_{33}(\tau^{(3)}_n,\tau^{(3)}_m).      \label{no31}
\eeq
Now, it is enough to show the correspondence of 
\eq{no30} to the rest $G^{\rm sym}_{23}$ and $G^{\rm sym}_{31}$. 
First, writing down $G^{(1)}_{13}(x_n,z)$ and $G^{(1)}_{13}(y_n,z)$ according 
to \eq{no22}, and then using \eq{no3} and \eq{no19}, we can rearrange them 
into similar forms as $G^{\rm sym}_{31}$ or $G^{\rm sym}_{23}$; for example,
\beq
G^{(1)}_{13}(y_n,z)=b-y_n+z-{1\over T( T_3+G_B(\tau_a,\tau_b))}
\left[T_3(b-y_n)^2+Tuz^2+T(1-u)(b-y_n+z)^2\right].    \label{no32}
\eeq
With the identification \eq{no21} and $z=\tau^{(3)}$, we get the following 
relations
\beq
 G^{(1)}_{13}(x,z)=G^{\rm sym}_{31}(\tau^{(3)},\tau^{(1)}),\label{no33}
\eeq
\beq
 G^{(1)}_{13}(y,z)=G^{\rm sym}_{23}(\tau^{(2)},\tau^{(3)}). \label{no34}
\eeq

In this paper, we have discussed the relation between different worldline 
Green functions $G^{(1)}_{ij}$ and $G^{\rm sym}_{ij}$. It is shown that 
both sets of these Green functions are transformed into each other under the 
relations \eq{no3}, \eq{no19} and \eq{no21}. Also the integral measures on 
the modular parameters $T_1$, $T_2$ and those on node points $\tau_a$, 
$\tau_b$ and $T$ are in correspondence to each other. From \eq{no3}, 
$\tau$ is related to $\tau^{(i)}$ as  
\beq
     \tau  =   (Tu-\tau^{(2)})\theta(Tu-\tau) 
                    +(\tau^{(1)}+Tu)\theta(\tau-Tu)        \label{no35}
\eeq
and this yields
\beq
     \int_0^T d\tau  =  \int_0^{Tu}d\tau^{(2)}
                       +\int_0^{T(1-u)}d\tau^{(1)},       \label{no36}
\eeq
which produces combinatorial factors in \eq{no11}, however such factors 
should be removed taking account of a constraint on proper times 
similarly to the analysis in \cite{holten}.

Our transformation rules are found through generalization of simple 
one-loop case. Our discussion on two-loop Green functions might 
be useful for other theories; for example, scalar and spinor QED diagrams to 
which super Green functions are applied \cite{SSqed} (see also \cite{sumino} 
where the same Green function as \eq{no15} and multi-loop diagrams are 
discussed in scalar QED). We finally anticipate that our analysis may be 
a help to clarification of multi-loop constructions in the worldline 
path integral approach.

\vspace{1cm}
\noindent
{\em Acknowledgments}

The author would like to thank J.W. van Holten, L. Magnea, K. Roland, 
M.G. Schmidt for useful suggestions and T. Onogi for discussions in the 
early stage of this work.

\newpage

%
\end{document}